\documentclass{iau} 
\usepackage{graphicx}
\usepackage{amsmath}
\newcommand{\solphys}{{\rm Solar Phys.}\ }
\newcommand{\apj}{    {\rm Astrophys. J.}\ }
\newcommand{\apjl}{   {\rm Astrophys. J. Lett.}\ }
\newcommand{\mnras}{  {\rm Mon. Not. Roy. Astron. Soc.}\ }
\newcommand{\jastp}{  {\rm J. Atmos. Sol. Terr. Phys.}}

\title[Solar oblateness \& asphericities variations] 
{Solar oblateness \& asphericities temporal variations: outstanding some unsolved issues}

\author[Jean Pierre Rozelot, Alexander Kosovichev \& Ali Kilcik]   
{Jean P. Rozelot$^1$,
Alexander G. Kosovichev$^2$
 \and Ali Kilcik$^3$}

\affiliation{$^1$Universit\'e C\^ote d'Azur, 77 Chemin des Basses Mouli\`eres, 06130 Grasse, France \\ email: {\tt jp.rozelot@orange.fr} \\
$^2$Center for Computational Heliophysics
and Department of Physics, New Jersey Institute of Technology, Newark, NJ 07102, USA
\\email: {\tt alexander.g.kosovichev@njit.edu}\\
$^3$ Akdeniz University Faculty of Science, Department of Space Science and
Technologies, 07058, Antalya, Turkey
\\email: {\tt alikilcik@akdeniz.edu.tr}}

\pubyear{2020}
\volume{354}  
\jname{Solar and Stellar Magnetic Fields: Origins and Manifestations}
\editors{A.G. Kosovichev, K. Strassmeier \& M. Jardine, eds.}

\begin{document}
	
\maketitle


\begin{abstract}
Solar oblateness has been the subject of several studies dating back to the nineteenth century. Despite difficulties, both theoretical and observational, tangible results have been achieved. However, variability of the solar oblateness with time is still poorly known. How the solar shape evolves with the solar cycle has been a challenging problem. Analysis of the helioseismic data, which are the most accurate measure of the solar structure up to now, leads to the determination of asphericity coefficients which have been found to change with time. We show here that by inverting even coefficients of f-mode oscillation frequency splitting to obtain the oblateness magnitude and its temporal dependence can be inferred. It is found that the oblateness variations lag the solar activity cycles by about 3 years.  A major change occurred between solar cycles 23 and 24 is that the oblateness was greater in cycle 24 despite the lower solar activity level.
Such results may help to better understand the near-subsurface layers as they strongly impacts the internal dynamics of the Sun and may induce  instabilities driving the transport of angular momentum. 
\end{abstract}
\keywords{Sun: heliosismology; Sun:activity; Sun: rotation; Sun: fundamental parameters.}

\section{Introduction}

The spherically symmetrical state represents a unique solution of the problem of hydrostatic equilibrium for a fluid mass at rest in the three-dimensional space. The problem complicates when the mass is rotating. For stars, the axial rotation modifies the shape of equilibrium by adding a centrifugal acceleration term to the total potential, breaking the spherical symmetry. The stellar sphere becomes an oblate figure, and we have no a priori knowledge of its stratification, the boundary shape, planes of symmetry, the  angular momentum transfer, etc. Moreover, when the rotation rate is not constant in radius and latitude, the surface deviates from a simple oblate figure, and it shape becomes more complicated, particularly, in the presence of internal stresses caused by magnetic fields, for instance.

Considering the Earth as a rotating ellipsoid in uniform rotation
$\omega$, Newton gave in 1687 for the first time, an approximate formulation of its flattening $f$, as a function of surface gravity $g_{s}$: 
$ f = \frac{5}{4}{\omega^2\cdot R_{eq}}/{g_{s}} $, where $R_{eq}$ is the equatorial radius.
Huyghens, in 1690, reformulated the flattening in the form $f = \frac{1}{2}{\omega^2\cdot R_{eq}}/{g_{s}}$, still commonly used as a first approximation.

Let us consider the case of a mass of polytropic gas of index $n$, rotating at a constant angular velocity $\omega$. The equilibrium configuration and shape of such a body is known since the works of \cite{Mil23} and \cite{Chan33}. By writing the mechanical equilibrium equations and seeking a solution in the form of a perturbed case of the non-rotating configuration, and neglecting high-order effects arising from $\omega^4$, and defining the boundary of the star by a constant null density level, 
the flattering is given by an equation of the type
$f = \upsilon{\omega^2}/{G\rho_c},$
where $G$ is the constant of gravitation, $\rho_c$ the density of the core, and $\upsilon$ is a term depending on the chosen polytropic index. Extensive computations can be found in \cite{Chan33}; for instance
for the solar case ($n$ = 3):
\begin{equation}
f = \left( 0.5 + 0.856 \frac{\rho_m}{\rho_c} \right) \frac{\omega^2 R_{eq}}{g_s}
\label{flattening}
\end{equation}
where $\rho_m/ \rho_c$, is the ratio of the mean to central density.
Even if such a formalism can be now considered as
outdated, it could be noticed that the approximation is still rather good for non polytropic structures with discontinuous variation of density, such as the Earth.

In the solar case, taking $\rho_c/ \rho_m$ = 107.168, $\omega$ = 2.85$\times$10$^{-6}$ rad/s, $R_\odot$ = 6.955080$\times$10$^{10}$ cm and $g$ = 2.74$\times$10$^{4}$ cm/s$^2$ (\cite{Allen2000}), it follows that $f=1.04\times10^{-5}$, in satisfying agreement with the best up-to-date determination of 8.55$\times$10$^{-6}$.

The story of the solar oblateness began in 1891 when Harzer (1891) introduced for the first time in a theory of solar rotation an oblateness of the Sun, estimating
$f$ as $\cong$ 6.32 $\times$ 10$^{-3}$. The history continued in 1895 when Newcomb (1895) described a rapidly
rotating solar interior in "such a way that the surfaces of equal density are non spherical". He demonstrated that if the difference between the equatorial and polar radii $\Delta r$ = $ R_{eq} - R_{pol}$ reached $\cong$ 500 mas, it would explain the discrepancy between the prediction of the Newtonian gravitational theory and the perihelion advance of Mercury observed by Le Verrier in 1859. However, measurements soon ruled out this hypothesis. The discrepancy between the observed advance of Mercury's
perihelion and the gravitational theory of planets was explained by the formalism developed by Einstein in 1905. 
In recent times, even though general relativity had given a satisfactory prediction of Mercury's perihelion, the argument was once again debated after Dicke's historical measurement of $\Delta r = 41.9 \pm3.3$ mas \cite{Dic70}. We know today that such measurements were inaccurate; nevertheless they have been a source of progress. Based on theoretical premises, Dicke (1970) proposed that the magnitude of the oblateness should be 8.1 $\times$ 10$^{-6}$, without any stress generated by other constraints (magnetic fields at first). Discussion of the historical data is certainly an interesting tour through different techniques. 
The precision required for determination of changes of the solar oblateness at the cutting edge of modern available techniques was set up, for instance, at the Pic du Midi observatory where a number of measurements was made (\cite{Roz11}, Table 2). But, even with a deconvolution of atmospheric effects, the measurements still suffered from atmospheric disturbances. The community was attentive to further progress coming from dedicated space experiments, first on balloon flights, and then on board  of spacecraft, mainly SoHO, SDO and in a lesser measure RHESSI (\cite{Fiv08}, \cite{Hud10}) and Picard (\cite{Irb19}). 
The main conclusions from this brief review have been summarized in \cite{Dam11}.

Through helioseismic measurements, considerable efforts have been made, at least since the eighties and up to now, in measuring from the odd-order frequency splitting coefficients the internal differential rotation of the Sun. Less progress has been made in analyzing the solar asphericity from the even-order frequency splitting measurements. Kuhn (1988) was the first to note that frequencies of solar oscillations vary systematically during the solar cycle, inferring the corresponding temperature change, but also noting that these variations could reflect changes in the solar structure due to variations of the Reynold's stresses or turbulent pressure. 
We try here to derive the global outer-limb shape temporal variations, assuming for this study that the site of the perturbation is very close to the surface.  

\section{Data}

Thanks to the Michelson Doppler Imager (MDI) 
(Scherrer et al., 1995) 
on Solar and Heliospheric Observatory (SoHO) and the Helioseismic and Magnetic Imager (HMI) 
(Scherrer et al., 2012) 
aboard NASA's Solar Dynamics Observatory (SDO), and their capability to observe with an unprecedented accuracy the surface gravity
oscillation (f) modes, it is possible to extract information concerning the coefficients of rotational frequency splitting, $a_n$. The odd $a_n$ coefficients ($n$ = 1, 3 ...) 
measure the differential rotation, whilst the even one ($n$ = 2, 4 ...) 
measure the degree of asphericity (i.e. departure from sphericity). The analysis was focused on the low-frequency medium-degree f-modes in the range of $\ell$ = 137-299, using the data covering nearly two solar cycles, from April 30, 1996, to June 4, 2017. The $a_{n}$ ($n$ even) coefficients) are a sensitive probe of the symmetrical (about the equator) part of distortion described by Legendre polynomials $P_{n}(\cos\theta)$.
Results published by 
Kosovichev and Rozelot (2018a,b)
showed that the asphericity of the Sun dramatically changes from the solar minimum to maximum. During the solar minimum (from 1996 to 1998) the asphericity was dominated by the $P_2$ and $P_4$ terms, while the $P_6$ contribution was negligible. It was shown that the ellipticity of the Sun is strongly affected during the solar cycle. We will try here to better quantify such temporal variations.    

According to the von Zeipel's theorem (1924), the solar-limb contours of temperature, density, or pressure should be nearly coincident near the photosphere. Rotation, magnetic fields, and turbulent pressure are the largest local acceleration sources that violate the von Zeipel's theorem \cite{Dic70}. Since (geometrical) asphericities are relatively small in the Sun, we may describe the distance from
the center, for instance, in terms of a constant isodensity level, (or, similarly, in terms of isotemperature or isogravity) by:
\begin{equation}
R(\cos\theta)|_{\rho={\rm constant}} = R_{sp} \left[ 1 + \sum_{n} c_n(R_{sp}) P_n (\cos\theta) \right]
\label{eq.rdep}
\end{equation}
where $R_{sp}$ is the mean limb contour radius, $\theta$ the angle to the symmetry axis (colatitude), and $P_n$ the Legendre polynomial of degree $n$. The asphericity is described by coefficients $c_n$, which are called quadrupole for $n$ = 2 ($c_2$) and hexadecapole for $n = 4$ ($c_4 $). Terms of higher orders are conventionally named by adding "-pole" to the degree number.
It is straightforward to determine $f$ from Eq.~\ref{eq.rdep} by means of the asphericities coefficients, $c_2$, $c_4$ and $c_6$, as $f = -\frac{3}{2}c_2 - \frac{5}{8}c_4 - \frac{21}{16}c_6 $. 

The measured splitting coefficients $a_n$ are related to the shape coefficients $c_n$ through a normalization factor $K$. An efficient method for calculating this factor was developed by Kuhn (1989) who showed that it was possible to invert the splitting data to obtain the structural asphericity; he obtained  
$a_n   = K c_n R_n (\ell)$.
Assuming $R_n$ = $R_{sp}$, as this analysis is conducted only very close to the surface (i.e. the seismic radius at the surface), the corresponding average factors are:
$$<a_2> = -0.546~ c_2 ~ R_{sp};~~~ <a_4> = 0.091 ~ c_4 ~ R_{sp};~~~ <a_6>= -1.274 ~ <c_6> ~ R_{sp} ~~~ $$
where the $a_n$ frequency splitting coefficients are measured in Hz (\cite{Kuh89}, to within /2$\pi$).

\section {Results}

\begin{figure}[t]
\centering
			\includegraphics[width=12.8cm,height=4.5cm]{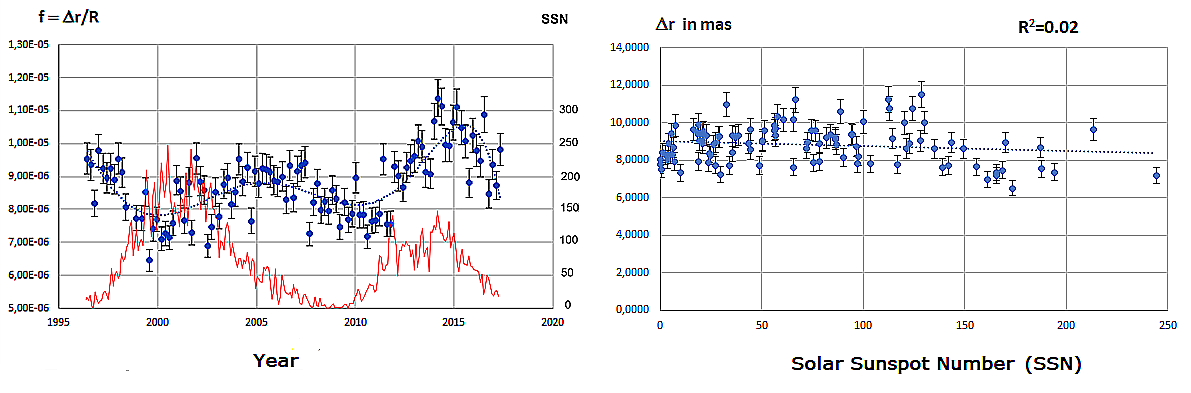}
     \caption{Left: Solar oblateness $f$ (left scale) and the solar sunspot number SSN (red, right scale) as a function of time. A periodic oscillation appears, with two minima around 2000 and 2011 and two maxima around 2005 and 2016. Right: The difference between the equatorial and polar radius $\Delta r$ (in mas) versus the solar activity described by the sunspot numbers. A slight anticorrelation is visible.}
         \label{fig1}
\end{figure}

\begin{figure}[t]
\centering
			\includegraphics[width=12.8cm,height=4.5cm]{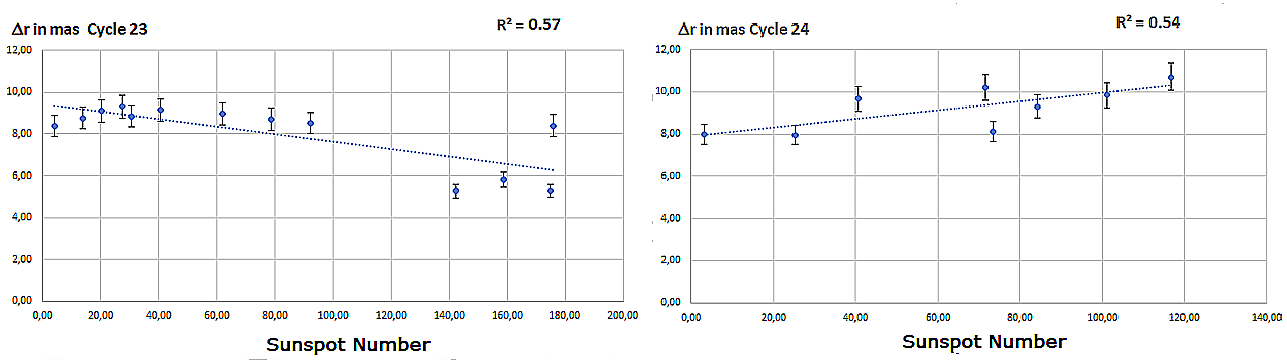}
     \caption{Annual mean difference between the equatorial and polar radius $\Delta r$ (in mas) versus the solar activity during cycles 23 (left) and 24 (right). The two cycles show a different behavior: a negative trend for cycle 23 and a positive one for cycle 24.
		(Source of the sunspot data: WDC-SILSO, Royal Observatory of Belgium, Brussels).}
         \label{fig2}
\end{figure}

Results displayed in Fig.~\ref{fig1}(left) show the solar oblateness $f$ as a function of time from 1996 to 2017. A periodic oscillation appears, with two minima around the years of 2000 and 2011 and two maxima around 2005 and 2016. If these minima and maxima correspond to the minima and maxima of the solar activity cycles, then there is a shift between the asphericity and activity of around 3 years. Fig.~\ref{fig1}(right) displays the difference between the equatorial and polar radii, $\Delta r$, in millisecond of arc (mas) versus  the solar activity, the sunspot number taken as a proxy \cite{Cle16}. In order to get a better view of the two cycles that are significantly different in the level of magnetic activity, we calculated variations of the $\Delta r$ annual means separately for these cycles.  
The two cycles show a different behavior as seen in Fig.~\ref{fig2}, left panel for cycle 23 and right panel for cycle 24. Straight lines show a linear regression fit. A negative trend for cycle 23 is noticeable, while a positive trend appears for cycle 24.

\section{Conclusion}

The analysis of the helioseismology data from the SoHO and SDO space missions permits to determine accurately the splitting rotational coefficients together with the structural shape parameters. 

The preliminary results obtained here by averaging the f mode frequency variations over two solar cycles for the whole observed angular degree range, $\ell$ = 137--299, lead to a mean solar oblateness of $f$ = 8.76 $\times 10^{-6}$. The deduced mean structural asphericity coefficients are respectively:    
$$c_2 = 8.08 \times 10^{-7};\;\; c_4 = -1.67 \times 10^{-5}	~~\mbox{\rm and~~} c_6 = 3.48 \times 10^{-7} $$
These even splitting coefficients vary in time as they depend on latitudinal inhomogeneities caused by aspherical perturbations due to the solar rotation,  magnetic fields beneath the surface, and even temperature variations.

It is shown that the solar oblateness is time dependent. However, its variation is quite complex, both in magnitude and time. If the  solar oblateness shows a periodicity of about 11 years, it does not follow exactly the solar cycle. Currently, we have no explanation for the $\sim 3$-year time lag of the flattering parameter $f$ relative to the activity cycle, bearing in mind that the $a_2$, $a_4$ and $a_6$ coefficients are respectively shifted from the solar cycle by around 0.1, 1.6 and -1.6 years (\cite{Koso18a}). The significant variations in time and the phase shifting according to the solar cycle activity (as seen in Fig.~\ref{fig1}) are probably two main reasons why the observational results from ground based instruments, balloon flights and satellite instruments seem to be inconsistent. An explanation has already put forward by \cite{Roz09} by considering the temporal variation caused by a change in the relative importance of the hexadecapolar and dipolar terms. At the time of high activity, only the dipolar moment $c_2$ has a significant effect, but at the time of low activity, $c_4$ is predominant; this results in a decrease of the total value of the oblateness.  Contribution of the $c_6$ term is less important due its low magnitude, but can be considered in a more detailed approach. Irbah et al. (2019) revisiting past solar oblateness measurements concluded that the solar oblateness "variations are in phase during odd cycles and anti-phase during even cycles", but the situation seems to be more complex. 
 
Clearly, we are very close to having the required data and boundary conditions to investigate deeper the solar shape structural coefficients and their changes during the two solar cycle that are significantly different in the level of magnetic activity.
Should the solar oblateness be determined accurately from space, this could help to disentangle the various contributions to the asphericity splittings of solar oscillation frequencies, and get insight into the physical processes that may be at play in the leptocline.\\

The work was partially supported by the NASA grants NNX14AB7CG and NNX17AE76A.\\


\begin{discussion}

\discuss{Krystof Helminiak}{About the radius variations -do we expect such variations in stars with stronger magnetic fields, like late type dwarfs? Could such variations be stronger and measurable?} 

\discuss{Jean Pierre Rozelot}{within the Sun, stronger magnetic fields lead to more important radius variability, particularly just below the surface (leptocline). We do expect the same for stars. Measurements are still difficult as we don't have yet accurate devices. This could be done in a next future. }

\end{discussion}

\

\end{document}